\documentclass[aip,apl,floatfix,showpacs,reprint,citeautoscript]{revtex4-1}

\usepackage{chemformula} 
\usepackage[T1]{fontenc} 
\usepackage{siunitx}
\usepackage{amssymb}
\usepackage{amsmath}
\usepackage{graphicx}
\usepackage{multirow,makecell}
\usepackage{booktabs}
\usepackage{footnote}
\usepackage[flushleft]{threeparttable}
\usepackage{comment}
\usepackage[toc,page]{appendix}

\usepackage{hyperref}
\hypersetup{
    colorlinks,%
    citecolor=blue,%
    linkcolor=blue,%
    urlcolor=blue
}

\providecommand{\U}[1]{\protect\rule{.1in}{.1in}}

\date{\today}

\begin{document}

\title{Nonradiative quenching of EPR signals in germanium-doped AlGaN: evidence for \textit{DX}-center formation}
\author{Jason Forbus}
\affiliation{Department of Physics, University of Alabama at Birmingham, Birmingham, Alabama 35233, USA}

\author{Darshana Wickramaratne}
\affiliation{Center for Computational Materials Science, US Naval Research Laboratory, Washington,
D.C. 20375, USA}

\author{John L. Lyons}
\affiliation{Center for Computational Materials Science, US Naval Research Laboratory, Washington,
D.C. 20375, USA}

\author{M.E. Zvanut}
\affiliation{Department of Physics, University of Alabama at Birmingham, Birmingham, Alabama 35233, USA}

\begin{abstract}
We present photo-electron paramagnetic resonance (EPR)
measurements and first-principles calculations that indicate germanium (Ge) is a \textit{DX} center in AlGaN.
Our photo-EPR measurements on Ge-doped AlGaN samples show no EPR spectra in the dark,
while persistent EPR spectra is observed upon photoexcitation with photon
energies greater than $\sim$1.3 eV.  Thermally annealing the samples decreased the EPR signal,
with the critical temperature to quench the EPR signal
being larger in the lower Al-content sample.
Using detailed first-principles calculations of Ge in AlGaN, we show all of these observations can be explained by accounting for the \textit{DX} configuration of Ge in AlGaN.
\end{abstract}

\maketitle

Aluminum nitride (AlN) and high-Al-content Al$_{x}$Ga$_{1-x}$N (AlGaN) alloys are key materials
for the future of high-power electronics \cite{wong2021ultrawide}, high-temperature electronics \cite{watson2015review} and ultra-violet optoelectronics \cite{amano2020}.
Enabling these technologies is predicated on the ability to achieve controlled doping \cite{lyons2024dopants}.
Germanium (Ge) has been explored as a potential $n$-type dopant,
motivated in part by experiments
that have demonstrated high solubility of Ge in GaN \cite{fireman2019high,ajay2016ge,kirste2013ge,fritze2012high}.

One challenge to $n$-type doping of ultra-wide band gap semiconductors
such as high-Al-content AlGaN alloys is compensation, either through point defects that act as acceptors or
the formation of \textit{DX} centers \cite{mooney1990deep,lyons2024dopants,gordon2014hybrid}.
A \textit{DX} center forms when a large lattice relaxation around a donor
dopant leads to the capture of two electrons.
This stabilizes the negative charge state with the neutral charge state
of the donor being metastable.
Hence, \textit{DX} centers inhibit efforts to achieve $n$-type doping, since the dopant effectively self-compensates
by establishing the Fermi level in the gap near the resulting (+/$-$) \textit{DX} level.
Prior first-principles calculations predicted a critical Al composition of 53\% at which Ge transitions from
being a shallow donor to a \textit{DX} center in AlGaN \cite{gordon2014hybrid}.

Several experimental studies have found evidence of carrier compensation in Ge-doped AlGaN, manifesting as a sharp drop in the
free carrier concentration.\cite{zhang1995growth,bansal2019effect,blasco2019electrical,bagheri2020nature,washiyama2021self,bagheri2021ge}
The origin of this observation has been debated: some studies have suggested this occurs due to compensation either by acceptor impurities such as carbon or cation vacancies \cite{washiyama2021self}, while other studies have suggested that Ge leads to a deep donor (+/0) level in the gap \cite{bagheri2020nature,bagheri2021ge}.
Establishing the properties of Ge in AlGaN, and answering the question of whether it is a \textit{DX} center or not
is especially pertinent, given that recent studies have also suggested that AlN could be $n$-type doped with Ge \cite{bagheri2023high}.

One possible reason for this ambiguity is that
transport measurements alone do not offer a direct way to deduce the stability of the
neutral charge state of the Ge donor (Ge$_{\rm cation}$; i.e., Ge occupying the cation site in AlGaN or AlN).
In this study we present a combination of photo-EPR studies and first-principles calculations that establish Ge$_{\rm Al}$ as a \textit{DX} center
in Al$_{x}$Ga$_{1-x}$N for $x$ greater than $\sim$0.5.
Our photo-EPR measurements identify evidence of the metastable neutral donor following photoexcitation.
This EPR signal remains persistent at low temperatures.
Performing \textit{in-situ} annealing shows the temperature to quench the EPR spectra
decreases as the Al content increases, contrary to what one expects for a deep donor in AlGaN.
Using detailed first-principles calculations we show all of these observations are consistent with Ge$_{\rm Al}$ being a \textit{DX} center.

Two Ge-doped Al$_{x}$Ga$_{1-x}$N samples
with $x$=0.65 and $x$=0.5, and a Si-doped sample with $x$=0.65, were used in this study.
Since our measurements are sensitive to the entire sample, we verified that the signal of interest originated
in the AlGaN film by studying a 3.4 $\mu$m
thick AlN film deposited on a sapphire substrate.  No response was observed
under any of the experimental conditions that we use in this study.
The AlGaN samples were grown by MOCVD on $c$-oriented sapphire \cite{bagheri2021pathway}.
Each of the samples had a 500 nm layer of
undoped AlN deposited directly on sapphire onto which 500 nm of Ge-doped or Si-doped AlGaN was grown.
The dopant concentration, measured by secondary ion mass spectrometry (SIMS) is listed in Table \ref{tab:sims}, along
with carrier concentrations obtained from Hall measurements.
\begin{table}[!htb]
\begin{center}
\caption{Aluminum content, $x$, in Al$_x$Ga$_{1-x}$N
and the dopant in square brackets,
dopant concentration measured by secondary ion mass
spectroscopy (SIMS), and electron concentration measured at room temperature
using Hall measurements reported in
Ref~\onlinecite{bagheri2021pathway}.}
\setlength{\tabcolsep}{6pt} 
\renewcommand{\arraystretch}{1.2} 
\begin{tabular}{ccc}
  \toprule\toprule
  {$x$ [Dopant]} & {Dopants (cm$^{-3}$)} & {Electrons (cm$^{-3}$)} \\
  \midrule
  65\% [Si]  & 1.0$\times$10$^{19}$ & 1.0$\times$10$^{19}$ \\
  65\% [Ge]  & 1.0$\times$10$^{19}$ & 2.7$\times$10$^{16}$ \\
  50\% [Ge]  & 3.0$\times$10$^{19}$ & 1.0$\times$10$^{18}$ \\
   \bottomrule\bottomrule
\end{tabular}
\label{tab:sims}
\end{center}
\end{table}

The EPR experiments were done with an $X$-band spectrometer at 9.4 GHz, which was set up for
low temperature (4-300 K) experiments using a liquid helium gas flow apparatus, cryostat,
and temperature controller. The samples were aligned with the $c$ axis perpendicular to the magnetic field,
to avoid any signal due to the sapphire substrate. The sample was illuminated with LEDs through slits
in the microwave cavity. These slits can be covered to ensure the sample was in darkness when necessary,
and the photon flux of the LEDs was controlled using neutral density filters.  The concentration of neutral donors was determined by calculating the double integral of the EPR spectrum and then comparing this value
to one obtained from a heavily phosphorus-doped Si powder reference.  To identify the neutral donor state, we determined the $g$-value
which was calculated using $g = C\frac{\nu}{B_0}$, where $\nu$ is the microwave frequency, B$_0$ is the magnetic field
where the EPR signal passes through zero, and $C$ is a correction factor found
using the phosphorus-doped Si powder which has a known $g$-value. Ideally, $C$=$\frac{h}{\mu_{B}}$, where $h$
is Planck’s constant (4.4$\times$10$^{-6}$ eV/GHz) and $\mu_{B}$ is the Bohr magneton (5.79$\times$10$^{-9}$ eV/GHz).
Measurement of the Si standard typically yields a value for $C$ of 715 Gauss/GHz.

{\it In-situ} annealing experiments were done for the Ge-doped samples. The sample was illuminated with an LED at 25 K
until the EPR signal saturated. The LED was then turned off and an initial EPR scan was taken.
The temperature was raised to a target temperature while the sample remained in the cavity;
the temperature was held constant for five minutes then lowered back to 25 K, where a subsequent EPR scan was taken.
The raising and lowering of the temperature was repeated until the EPR signal was thermally quenched.

To interpret these results we perform first-principles calculations
based on density functional theory (DFT) \cite{HK64,KohnPR1965}, using the projector-augmented wave (PAW) potentials \cite{BlochlPRB1994}
as implemented in the Vienna Ab-initio Simulation Package (VASP) code \cite{KressePRB1996,kresse1994ab}.
We use the Heyd-Scuseria-Ernzerhof (HSE) hybrid functional \cite{HeydJCP2003,*HeydJCP2006}
for all of our calculations.  The energy cutoff for the plane-wave basis set is 500 eV.  The fraction of nonlocal Hartree-Fock exchange
is set to 0.33 for AlN; this results in band gaps and lattice parameters that agree with the experimental values.
Defect formation energies and thermodynamic transition levels are calculated using the standard
supercell approach \cite{FreysoldtRMP2014} with 96-atom supercells.  The lattice parameters of the supercell
are held fixed while the atomic coordinates are relaxed until the  forces are below 5 meV/Angstrom.

To describe the electronic structure of the alloy, the lattice parameters are scaled in accordance with
Vegard's law to match the lattice parameters for a given Al content in AlGaN.
This approach has previously been successfully applied to describe defects
in indium containing GaN alloys \cite{dreyer2016gallium,wickramaratne2020deep}.
The results are comparable to supercell calculations where Al and Ga atoms are explicitly included into a 96-atom supercell
to match a given alloy composition, and to virtual crystal approximation (VCA) calculations of Ge in AlGaN (see Supplementary Material)
The formation energies and charge-state transition levels of Ge are determined
using a single ($\frac{1}{4}$,$\frac{1}{4}$,$\frac{1}{4}$) $k$-point.  Spin polarization is included
for all of our calculations.  To analyze the nonradiative and optical properties we
construct one-dimensional configuration coordinate diagrams using the Nonrad code \cite{turiansky2021nonrad}.

\begin{figure}[!htb]
\includegraphics[width=8.5cm]{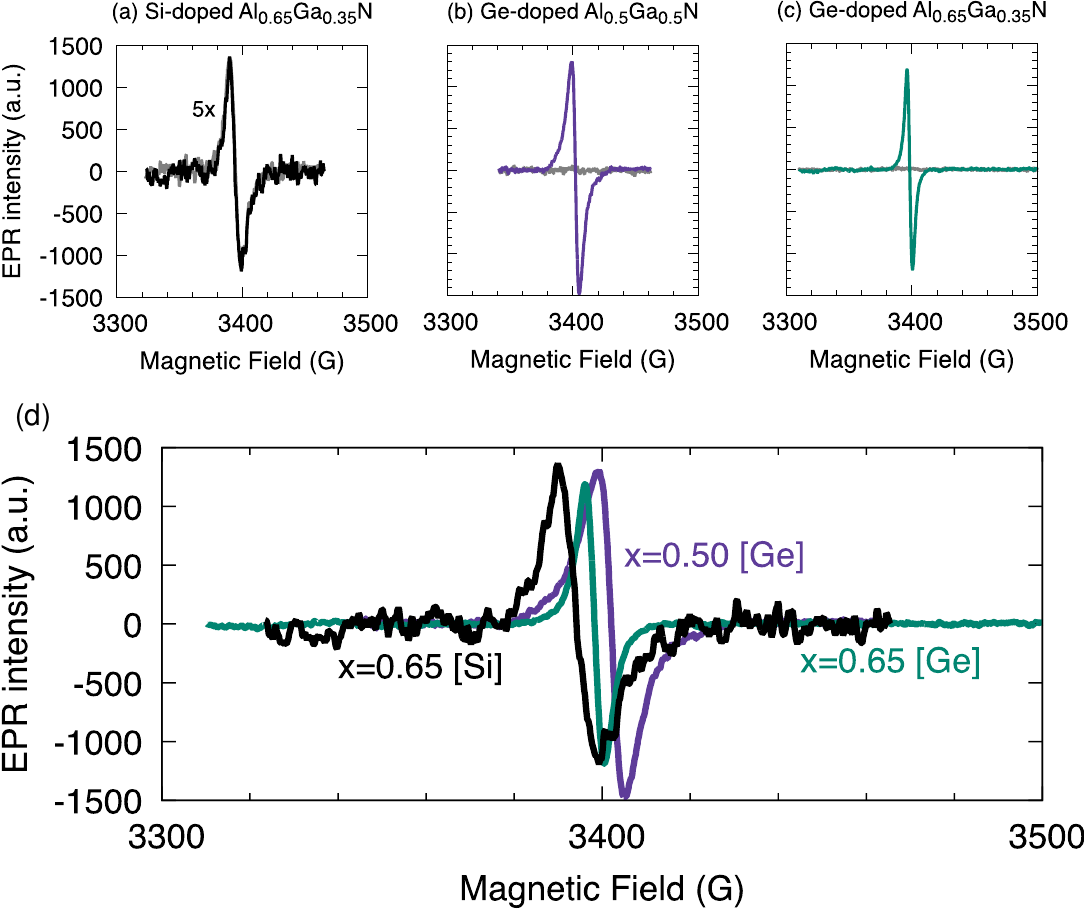}
\caption{
EPR spectra measured in the dark (grey) and post-illumination in Si and Ge-doped Al$_{x}$Ga$_{1-x}$N. (a) Si-doped Al$_x$Ga$_{1-x}$N
for $x$=0.65 (multiplied by a factor of 5 for clarity), (b) Ge-doped Al$_x$Ga$_{1-x}$N for $x$=0.5 and (c) Ge-doped Al$_x$Ga$_{1-x}$N for $x$=0.65.
All of the measurements are taken at 25 K. (d) EPR spectra from (a-c) obtained post illumination.
}
\label{fig:HF}%
\end{figure}

Figure \ref{fig:HF} illustrates the EPR signal measured at 25 K for the three samples prior to and after illumination.
For the Si-doped sample we observed an EPR signal in the dark, and the spectrum
was unaffected by subsequent illumination, as expected with Si being a shallow donor in AlGaN with 65\% Al.
For the Ge-doped samples we did not observe any EPR signal above our detection limit of 10$^{15}$ cm$^{-3}$ in the dark.
The threshold to photoexcite the EPR signal was 1.3 eV in the Ge-doped samples.  The spectra shown in
Fig.~\ref{fig:HF} were obtained using larger photon energies so that a clearer signal could be presented.
The EPR signal in the Ge-doped samples also persisted for at least an hour post-illumination.

The photo-induced EPR spectra are replotted in Fig.~\ref{fig:HF}(d) to emphasize the different $g$-values.
The EPR signal in all three samples corresponds to an isotropic $g$-value of $\sim$1.98.
Notably, the $g$-value for the spectra in the Si-doped sample with 65\%~Al differs from that in the 
Ge-doped sample of similar Al content.  This difference is possibly due to the different local strain
induced by each dopant \cite{bansal2019effect}.
The magnitude of the measured $g$-values and the trend of increasing $g$-values
with increasing Al content (Table \ref{tab:gvalue} and Fig.~\ref{fig:HF}(d)) are in agreement with
prior studies on AlGaN alloys \cite{bayerl2001g}.
\begin{table}[!htb]
\begin{center}
\caption{Experimentally determined $g$-values from our measurements on Si and Ge-doped AlGaN compared against results in
Ref~\onlinecite{bayerl2001g}.}
\setlength{\tabcolsep}{6pt} 
\renewcommand{\arraystretch}{1.4} 
\begin{tabular}{lccc}
  \toprule\toprule
  {Al} & Dopant & {$g$} & {Reference}  \\
  \midrule
  65\% & Si & 1.9791(1) & Current work \\
  50\% & Ge & 1.9749(3) & Current work \\
  65\% & Ge & 1.9777(1) & Current work \\
  52\% & Si & 1.9733 & Ref.~\onlinecite{bayerl2001g} \\
  75\% & Si & 1.9802 & Ref.~\onlinecite{bayerl2001g} \\

   \bottomrule\bottomrule
\end{tabular}
\label{tab:gvalue}
\end{center}
\end{table}
The Ge-doped Al$_{0.5}$Ga$_{0.5}$N sample had the largest zero-crossing which results in the lowest $g$-value of all three samples.
The linewidths of the three samples ranged from 5-8 G at 25 K.
Increasing temperature led to linewidth narrowing.
The signals also did not saturate with microwave power up to 10 mW.
Each of these characteristics are consistent with the EPR spectra arising from neutral donors \cite{carlos1993electron}.

\begin{figure}[!htb]
\includegraphics[width=7.5cm]{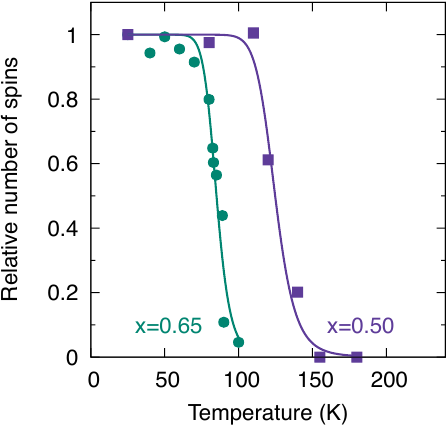}
\caption{
Relative number of spins obtained from EPR measurements of Ge-doped Al$_{x}$Ga$_{1-x}$N for $x$=0.65 (green circles)
and $x$=0.50 (purple squares) measured at 25 K after being annealed at specified temperature for five minutes.
The size of the symbols is larger than any uncertainty in the temperature.  The solid lines are a fit to extract
the effective capture barriers.  See main text for further details on the fitting.
}
\label{fig:epr_temp}%
\end{figure}

For the two Ge-doped samples where the EPR spectra persisted in the dark we conducted {\it in-situ}
annealing experiments and monitored the change in the EPR spectra.
In Fig.~\ref{fig:epr_temp}, the change in spin density after annealing for five minutes at a given temperature is shown relative
to the initial spin density prior to annealing experiments.
The initial spin density in the 50\% Al sample is 2.3$\times$10$^{17}$ cm$^{-3}$
and it is 4.8$\times$10$^{16}$ cm$^{-3}$ in the 65\% Al sample. The number of spins in both samples
remains fairly constant for low anneal temperatures beyond which the number rapidly
decreases to a point where the EPR signal disappears and no spins can be detected.
For the 50\% AlGaN sample, the EPR signal is completely quenched at 160 K, while it is quenched at 100 K for the
sample with 65\% Al.

Our experiments on the Ge-doped AlGaN samples lead to three principal observations
that require a microscopic explanation: the lack of an EPR signal in the Ge-doped samples when measured in the dark,
the observation of a neutral donor EPR signal only upon photoexcitation with energies greater than 1.3 eV,
and finally the quenching temperature of the EPR signal decreasing with increasing Al content.
Using first-principles calculations we highlight how each of these
observations are traits of \textit{DX} centers \cite{mooney1990deep}.

\begin{figure}[!htb]
\includegraphics[width=7.5cm]{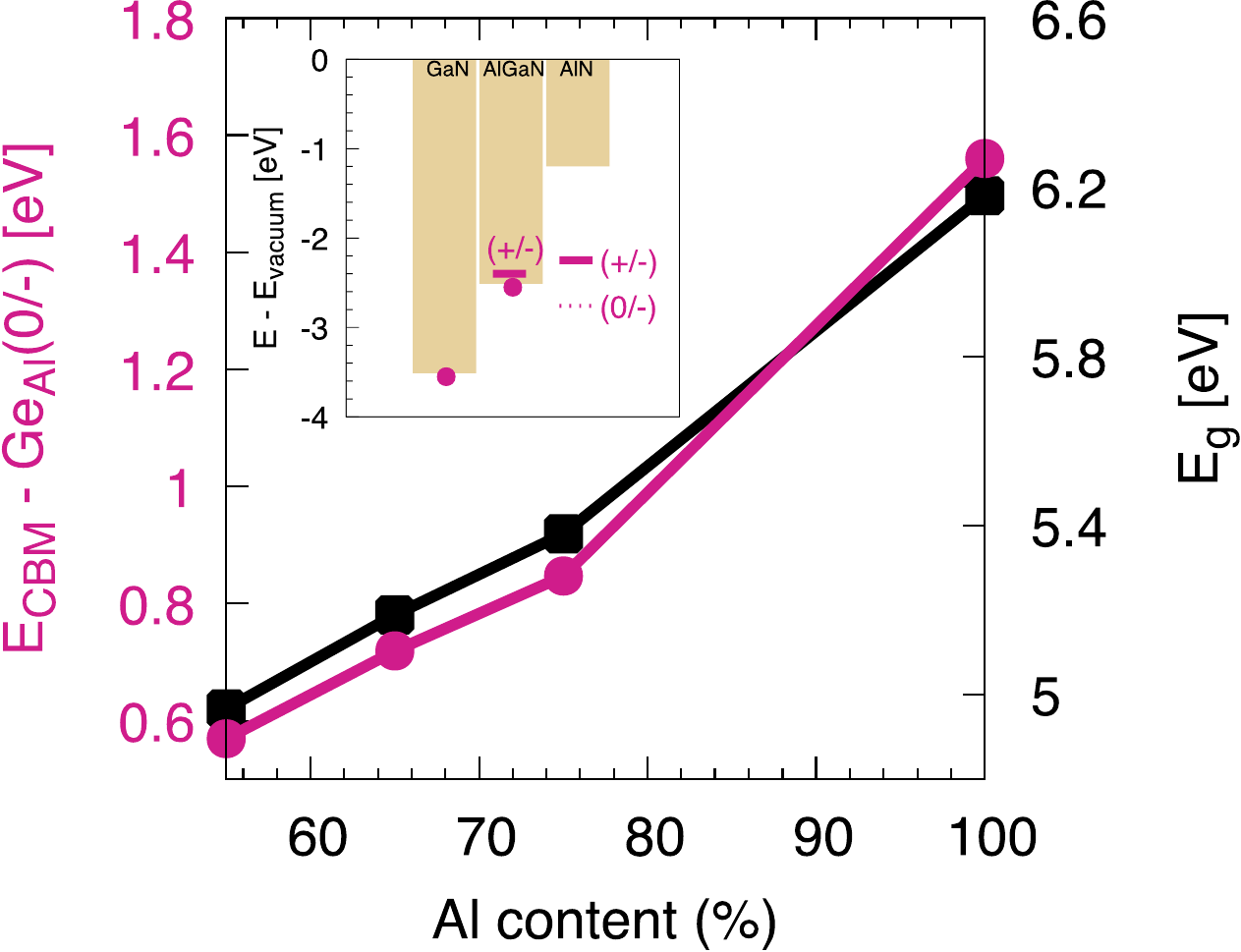}
\caption{
 Energy difference between the Ge$_{\rm cation}$ (0/$-$) level and the AlGaN CBM (magenta-circles, left vertical axis) and the as calculated AlGaN band gap ((black-squares, right vertical axis) plotted as a function of Al content.
The inset illustrates the (+/$-$) and (0/$-$) level of Ge in AlN and AlGaN with respect to the conduction band minimum.
In the 50\%~AlGaN alloy and in GaN, Ge is a shallow donor which we denote using the magenta colored circle.
}
\label{fig:algan}%
\end{figure}

Consistent with previous first-principles studies  \cite{gordon2014hybrid} we find
Ge substituting on the Al site is a \textit{DX} center.
The (+/$-$) level of Ge$_{\rm Al}$ is 1.06 eV below the AlN CBM, while the metastable (0/$-$) level is 1.57 eV below
the AlN CBM, as illustrated within the inset of Fig.~\ref{fig:algan}.
In AlN, the metastable neutral charge state, Ge$_{\rm Al}^{0}$, leads to a breathing-like relaxation of the N ions
that are nearest-neighbor to Ge$_{\rm Al}$, resulting in Ge$-$N bonds
that are equivalent in length.  In the \textit{DX} state, Ge$_{\rm Al}^{-}$,
there is a large relaxation of the axial N away from Ge, while the three basal plane Ge-N bond lengths are equivalent in length.
We calculated the formation energies and thermodynamic transition levels for Ge substituting on the cation site in AlGaN
for $x$=0.50, 0.55, 0.65 and 0.75 using strained supercell calculations.

For $x$=0.50, Ge$_{\rm cation}$ acts as a shallow donor and the (+/$-$) level is
above the AlGaN CBM.  For $x$=0.55, 0.65 and 0.75, Ge is stable as a \textit{DX} center with (+/$-$) levels that are below the AlGaN CBM.
The energy difference between the (+/$-$) level and the AlGaN CBM is 0.07 eV ($x$=0.55), 0.22 eV ($x$=0.65) and 0.35 eV ($x$=0.75)
for these intermediate Al compositions.
The energy difference between the metastable (0/$-$) level
and the AlGaN CBM also increases with Al content (Fig.~\ref{fig:algan}).
The atomic relaxations for each of these alloy compositions are similar
to the case of Ge$_{\rm Al}$ in AlN; we find a breathing-like relaxation for Ge$_{\rm cation}^{0}$ and Ge$_{\rm cation}^{+}$, while for Ge$_{\rm cation}^{-}$ we find a large
outward displacement of the axial N.

\begin{figure*}[!htb]
\includegraphics[width=17.5cm]{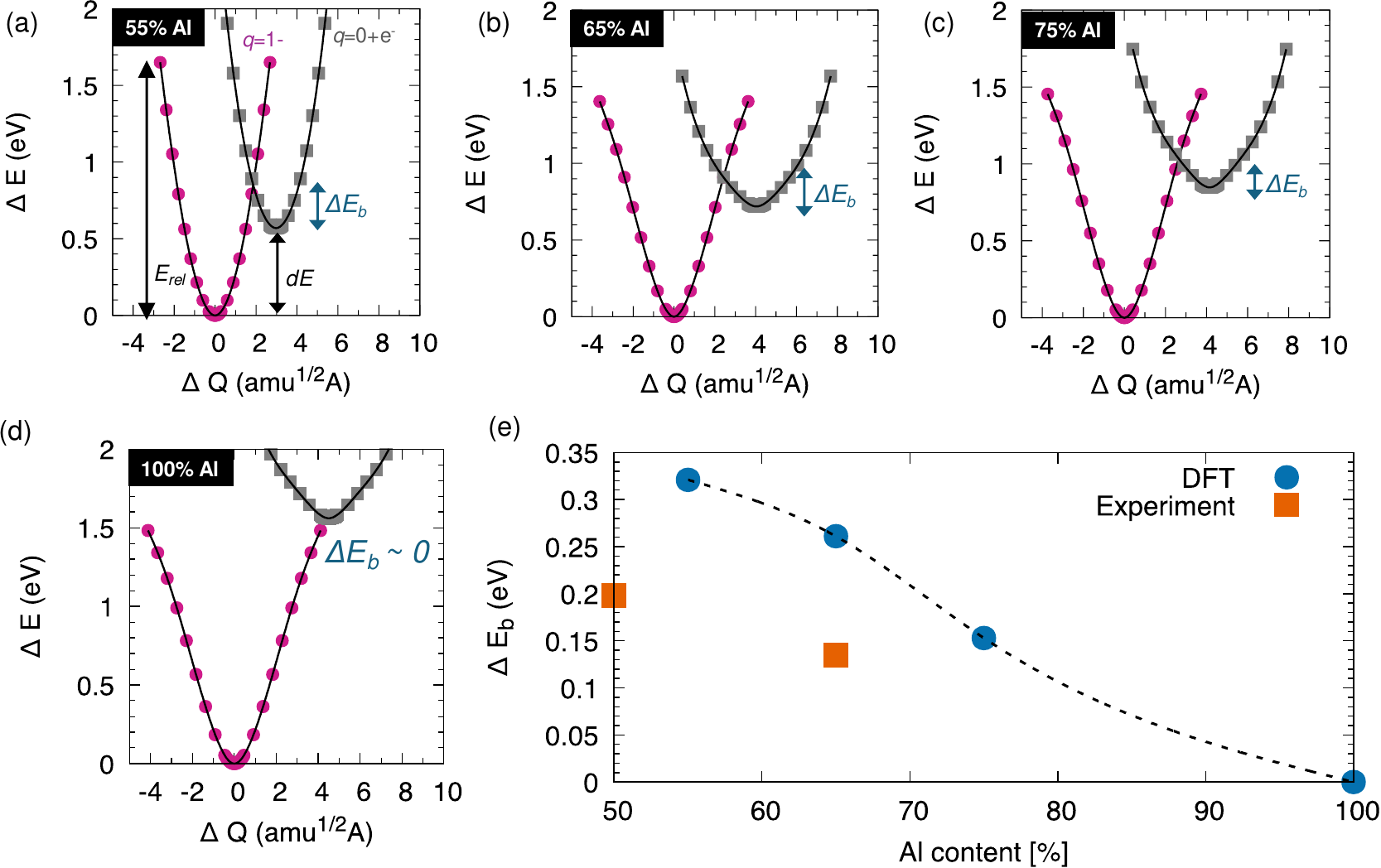}
\caption{
Configuration coordinate diagrams for Ge$_{\rm cation}$ in AlGaN alloys with (a) 55\%~Al, (b) 65\%~Al, (c) 75\%~Al, and (d) 100\%~Al.
The relaxation energy of the initial state, $E_{\rm rel}$ and the ionization energy with respect to the AlGaN CBM, $dE$ is illustrated in (a).
The values for $dE$ are from Fig.~\ref{fig:algan}.
The classical capture barrier, $\Delta E_b$ for each composition is illustrated with a blue vertical arrow.  (e)  Change in $\Delta E_b$ as a function of Al content obtained from panels (a-d) plotted with blue circles.  The effective capture barrier by fitting Fig~\ref{fig:epr_temp} is illustrated with orange squares.  See main text for details on the fitting.
}
\label{fig:ccd}%
\end{figure*}

The EPR-active Ge$_{\rm cation}^{0}$ state is metastable for $x$=0.55, 0.65 and 0.75, and 1 in our calculations, explaining why we
do not observe any EPR spectra in the dark (Fig.~\ref{fig:HF}(b-c)) in the Ge-doped samples.
Photoionization of the \textit{DX} center above a critical photon energy
would convert the EPR-silent Ge$_{\rm cation}^{-}$ to Ge$_{\rm cation}^{0}$ state and activate the EPR signal.
We calculate configuration-coordinate diagrams
to describe this optical absorption process for
$x$=0.55, 0.65, 0.75 and 1.0 (Fig.~\ref{fig:ccd}(a-d)).
The peak optical absorption energy based on the Franck-Condon approximation
ranges from $\sim$1.5 eV (for $x$=0.55, 0.65, 0.75)
to $\sim$2.5 eV ($x$=1).  The large lattice relaxation between the negative
to the neutral charge state leads to peak optical absorption energies that are significantly larger than $dE$ (Fig.~\ref{fig:algan}).
The large peak optical absorption energy (relative to $dE$) is consistent with our experiments where we find photon energies
greater than $\sim$1.3 eV are required to
photoionize the negative charge state of Ge$_{\rm cation}$ and activate the EPR signal.

Our calculated configuration-coordinate diagrams makes clear that for the $x$=0.55, 0.65 and 0.75 alloys there is a finite barrier, $\Delta E_b$, to convert
from the EPR active Ge$_{\rm cation}^{0}$ state back to the EPR silent Ge$_{\rm cation}^{-}$ state.
Because we perform our EPR measurements at low temperature (25 K), there is insufficient thermal energy post-illumination for
a nonradiative capture process to occur, explaining why the EPR signal remains persistent.

It might be tempting to interpret the subsequent thermal quenching of the EPR signal as
the ionization of the Ge$_{\rm cation}$ level.
Within this interpretation,
the difference in temperature required to quench the EPR signal as the Al content changes
may appear to correspond to changes in the depth of the Ge level with respect to the CBM.
However, $dE$ increases with Al content (Fig.~\ref{fig:algan}), which
is at odds with the higher Al content sample requiring a lower anneal temperature.
Rather, we propose this trend in quenching temperatures as a function of Al content corresponds to changes in the nonradiative capture barrier that converts the neutral to the negative charge state.
In Fig.~\ref{fig:ccd}(e), we illustrate $\Delta E_b$ as a function of Al content obtained from our first-principles calculations of the
configuration-coordinate diagrams.

To compare the calculated $\Delta E_b$ with our EPR quenching measurements,
we interpret the thermal quenching of the relative number of spins as nonradiative
capture of the photo-excited electrons in the CBM into the Ge$_{\rm cation}$ (0/$-$) level.
We fit our temperature-dependent EPR data (Fig.~\ref{fig:epr_temp}) to a Mott-Seitz type expression
for nonradiative capture \cite{di2013advances}, $\frac{n(T)}{n} = \frac{1}{1+\alpha e^{[-E_0/k_BT]}}$,
where $n$ is the number of spins at the lowest temperature (25 K) that we perform our measurements,
$\alpha$ is a frequency factor, and $E_0$ is the experimental effective capture barrier for the nonradiative process.
Keeping $\alpha$ fixed at 10$^8$ we find
$E_0$ = 0.198 eV for $x$=0.5 and $E_0$ = 0.135 eV for $x$=0.65, which is illustrated in Fig.~\ref{fig:epr_temp} and Fig.~\ref{fig:ccd}(e).
If we allow $\alpha$ and $E_0$ to vary in our fitting, we find
$\alpha$=2$\times$10$^7$ and $E_0$ = 0.182 eV for $x$=0.5, and
$\alpha$=5$\times$10$^8$ and $E_0$ = 0.147 eV for $x$=0.65.
Consistent with our calculations, we find the capture barrier extracted from our fitting decreases as the Al content increases.

It might be surprising to note that $\Delta E_b$ decreases as the Al content increases, since
one may assume that $\Delta E_b$ increases with $dE$.
However, If $E_{\rm rel}$ is greater than $dE$, $\Delta E_b$ decreases as $dE$ increases.
This is apparent if we consider an analytical expression for $\Delta E_b$ as $\frac{(dE - E_{\rm rel})^2}{4E_{\rm rel}}$,
assuming two identical harmonic one-dimensional configuration-coordinate diagrams \cite{wickramaratne2018comment}.
Our calculations of the configuration-coordinate diagrams as a function of Al content (Figs.~\ref{fig:ccd}(a-d))
clarify that $E_{\rm rel}$ for the negative charge state is always large compared to $dE$.
The large relaxation energy is due in part to the lattice relaxation associated with the \textit{DX} state,
explaining why the quenching temperature of the EPR signal decreases for larger Al content.

The experimental and theoretical results presented in this study provide clear evidence that Ge in AlGaN acts as a \textit{DX} center for Al contents
greater than $\sim$50\%.  Our photo-EPR experiments exhibit two traits of Ge$_{\rm Al}$ being a \textit{DX} center: the absence of an EPR signal in the dark, consistent with the neutral charge state of Ge being metastable, and also photoexcitation activating the neutral donor EPR signal that remains persistent at low temperature.
Heating the sample quenches the persistent EPR signal.
Surprisingly, we find the temperature to quench the EPR signal decreases with
increasing Al content.
Using detailed first-principles calculations we show this difference in temperature
corresponds to a reduction in the magnitude of the capture barrier
associated with the nonradiative process in the EPR thermal quenching measurement.
The increase in anneal temperature with deceasing Al content seen in Fig.~\ref{fig:epr_temp}
is a consequence of the large lattice relaxation associated with
Ge$_{\rm cation}$ being a \textit{DX} center.

See the supplementary material for additional details on the 
alloy calculations.

\begin{acknowledgements}
The work at UAB was supported as part of the Ultra
Materials for a Resilient Energy Grid, an Energy Frontier
Research Center funded by the U.S. Department of Energy,
Office of Science, Basic Energy Sciences under Award No.
DE-SC0021230
D.W and J.L.L were supported by the Office of Naval Research through the Naval Research Laboratory's Basic Research Program.
The calculations were supported in part by the DoD Major Shared Resource Center at AFRL.
\end{acknowledgements}

\section*{Data Availability}
The data that supports the findings of this study are available from the corresponding author upon reasonable request.


%

\end{document}